\documentclass[11pt,dvips]{article}
\usepackage{graphicx}
\usepackage{amsmath}
\usepackage{amssymb}
\usepackage{latexsym}
\usepackage{color}
\begin{document}
\thispagestyle{empty}
\begin{center}

\vspace{1.8cm}

 {\bf Geometric Measure of Pairwise Quantum Discord for Superpositions of Multipartite Generalized  Coherent States}\\

\vspace{1.5cm}

{\bf M. Daoud}$^{a}${\footnote { email: {\sf
m$_{-}$daoud@hotmail.com}}} and {\bf R. Ahl Laamara}$^{b,c}$
{\footnote { email: {\sf ahllaamara@gmail.com}}}

\vspace{0.5cm}
$^{a}${\it Department of Physics, Faculty of Sciences, University Ibnou Zohr,\\
 Agadir,
Morocco}\\[1em]

$^{b}${\it LPHE-Modeling and Simulation, Faculty  of Sciences,
University
Mohammed V,\\ Rabat, Morocco}\\[1em]

$^{c}${\it Centre of Physics and Mathematics,
CPM, CNESTEN,\\ Rabat, Morocco}\\[1em]

\vspace{3cm} {\bf Abstract}
\end{center}
\baselineskip=18pt
\medskip
We give the explicit expressions of the pairwise quantum
correlations present in superpositions of multipartite coherent
states. A special attention is devoted to the evaluation of the
geometric quantum discord.  The dynamics of quantum correlations
under a dephasing channel is analyzed. A comparison of geometric
measure of quantum discord with that of concurrence shows that
quantum discord in multipartite coherent states is more resilient to
dissipative environments than is quantum entanglement. To illustrate
our results, we consider some special superpositions of
Weyl-Heisenberg, $SU(2)$ and $SU(1,1)$ coherent states which
interpolate between Werner and Greenberger-Horne-Zeilinger states.

\newpage

\section{Introduction}

The total correlation in quantum states of a composite system can be
split into a classical and a quantum parts. Entanglement is a kind
of quantum correlation without classical counterpart. It has
triggered off many efforts for a deeper understanding of the
difference between classical and quantum correlations. Nowadays, it
is well established that entanglement is not only a concept in
quantum physics but also a fundamental resource for quantum
information processing and necessary for some quantum information
tasks like quantum teleportation, quantum cryptography and universal
quantum computing (for review, see
\cite{NC-QIQC-2000,Vedral-RMP-2002,Horodecki-RMP-2009, Guhne}).
However, some recent investigations showed that entanglement is only
a one special kind of quantum correlations. Indeed, unentangled
quantum states can also possess quantum correlations which play a
relevant role in improving quantum communication and information
protocols better than their classical counterparts
\cite{Knill,Datta-PRL100-2008}. Therefore, as the non-classicality
of correlations present in bipartite and multipartite quantum states
is not due solely to the presence of entanglement, there was a need
of a measure to characterize and quantify the non-classicality — or
quantumness — of correlations which goes beyond entanglement. A
recent catalogue of measures for non classical correlations is
presented in \cite{vedral}. The most popular among them is the
so-called quantum discord introduced by Henderson and Vedral
\cite{Vedral-et-al} and (independently) Ollivier and Zurek
\cite{Ollivier-PRL88-2001} who showed that when entanglement is
subtracted from total quantum correlation, there remain correlations
that are not entirely classical of origin. Now, it is commonly
accepted that the most promising candidate to measure quantum
correlations is quantum discord.  It has attracted considerable
attention and continues to be intensively investigated in many
contexts \cite{vedral}. The quantum discord is defined as the
difference between quantum mutual information and classical
correlation in a bipartite system
\cite{Vedral-et-al,Ollivier-PRL88-2001}. It has been calculated
explicitly only for a rather limited set of two-qubit quantum states
and expressions for more general quantum states are still not known.
This is essentially due to the fact that the evaluation of quantum
discord involves an optimization procedure which is in general a
difficult task to perform. To overcome this difficulty a geometric
measure for quantum discord was proposed recently \cite{Dakic2010}.
This is defined, by means of Hilbert-Schmidt norm, as the nearest
distance between the quantum state under consideration and the
zero-quantum discord states. The explicit expressions of the
geometric quantum discord have been obtained only in some few cases
including Gaussian states \cite{Adesso1,Giorda} and superpositions of Dicke states \cite{Yin}. \\

In this paper we shall be mainly concerned with the geometric
measure of quantum discord of generalized coherent states. The main
reason is that the coherent states constitute a special instance of
non orthogonal states whose entanglement properties have received a
special attention during the last years (for a recent review see
\cite{Sanders} and references therein). In this respect, it is
important to investigate the quantum correlations in such quantum
states beyond entanglement. Also, the coherent states are of
paramount importance in physics (e.g., in quantum optics and quantum
information theory) and mathematical physics (e.g., in probability
theory, applied group theory, path integral formalism and theory of
analytic functions) \cite{Klauder,Perelomov,Ali,Gazeau}.\\

The paper is organized as follows. In order to study the pairwise
quantum correlations, we present in Section 2 two different schemes
of bipartite partitioning and qubit mapping of a balanced
superposition of multipartite coherent states. Section 3 is devoted
to the explicit derivation of the geometric measure of quantum
discord. The dynamical evolution of bipartite quantum correlations
(entanglement and quantum discord) under a dephasing channel is
considered in Section 4. Finally, as illustration, some special
cases are considered in the last section. Concluding remarks close
this paper.

\section{ Superpositions of multipartite coherent states, bipartite partitioning and qubit mapping}

\subsection{ Superpositions of multipartite coherent states}
We begin by recalling  some elements of the Perelomov group
theoretic procedure to construct  coherent states for a quantum
system whose  dynamical symmetry is described by a Lie group $G$
(connected and simply connected, with finite dimension). Let $T$ be
a unitary irreducible representation of $G$ acting on  the Hilbert
space of the system. The construction of Perelomov coherent states
requires a specific choice of the reference (the ground) state. A
coherent state $|\Omega\rangle$ is determined by a point $\Omega$ in
the coset space $G/H$
 where the isotropy subgroup $H \subset G$ consists of all the group
elements that leave the reference state invariant. The dimension of
the coset space $G/H$ determines the number of the complex variables
labeling the Perelomov coherent states. A very important property is
the identity resolution in terms of the coherent states:
\begin{equation}
\int_{G/H} d \mu(\Omega) |\Omega\rangle\langle\Omega| = I
\label{mesure}
\end{equation}
where $d \mu(\Omega)$ is the invariant integration measure on $G/H$
and the integration is over the whole manifold $G/H$  and $I$ is the
identity operator on the Hilbert space. The choice of the reference
state leads to systems consisting of states with properties closest
to those of classical states \cite{Perelomov}.  As special examples
one may quote  the Glauber coherent states associated with the
Weyl-Heisenberg group $H_{3}$ which are defined on the complex plane
${\Bbb{C}} =  H_3 /  U(1)$, the spin coherent states  defined on the
unit sphere ${\Bbb{CP}}^1 = SU(2) / U(1)$ and $SU(1,1)$ coherent
states  defined on unit disc ${\Bbb{CB}}^1 = SU(1,1) / U(1)$. All
these coherent states are labeled by a single complex variable.

For a composite system of $n$ noninteracting quantum subsystems,
 the Hilbert space  is  the tensor product of $n$ copies of
single particle Hilbert space
$$ {\cal H} = {\cal H}_1 \otimes {\cal H}_2\otimes \cdots \otimes {\cal H}_n.$$
The coherent states are not orthogonal to each other with respect to
the positive measure in (\ref{mesure}). Thus, the over-completion
relation makes possible the expansion of an arbitrary state of the
Hilbert space ${\cal H}$ in terms of coherent states of the quantum
system under consideration. It follows that when one has a
collection $n$ of particles or modes, the whole Hilbert space is a
tensorial product and any multipartite state can be written as a
superposition of tensorial product of the coherent states $\vert
\Omega_1 \rangle\otimes\vert \Omega_2 \rangle\cdots \otimes\vert
\Omega_n \rangle \equiv \vert \Omega_1, \Omega_2, \cdots \Omega_n
\rangle$. Indeed, using the resolution to identity,  any state
$\vert \psi \rangle $ in ${\cal H}$ can be expanded as
\begin{equation}
 \vert \psi \rangle  = \int ~ d\mu(\Omega"_1) ~ d\mu(\Omega"_2)\cdots ~ d\mu(\Omega"_n)\vert \Omega"_1, \Omega"_2, \cdots \Omega"_n  \rangle
 \langle  \Omega"_1, \Omega"_2, \cdots \Omega"_n
 \vert \psi \rangle \label{ncs}
\end{equation}
reflecting that any multipartite state can be viewed as a
superposition of product coherent states. The multipartite  state
(\ref{ncs}) can be reduced to a sum if the function
$$ \psi (\Omega"_1, \Omega"_2, \cdots \Omega"_n) = \langle
\Omega"_1, \Omega"_2, \cdots \Omega"_n
 \vert \psi \rangle$$
   is expressed as a sum of delta functions.
To simplify our purpose, we set
$$ \psi (\Omega"_1, \Omega"_2, \cdots \Omega"_n) =
\prod_{i = 1}^n \delta(\Omega_i - \Omega"_i) + e^{im \pi} \prod_{i =
1}^n \delta(\Omega'_i - \Omega"_i).$$ This gives the following
equally weighted or balanced superpositions of multipartite coherent
states
\begin{equation}
\vert \psi \rangle \equiv \vert \Omega, \Omega', m, n \rangle ={\cal
N}(\vert \Omega_1 \rangle\otimes \vert \Omega_2\rangle\otimes
\cdots\otimes \vert \Omega_n \rangle
+e^{im\pi}|\Omega'_1\rangle\otimes |\Omega'_2 \rangle\otimes
\cdots\otimes |\Omega'_n \rangle) \label{multi-cs}
\end{equation}
where $m \in \mathbb{Z}$ and ${\cal N}$ is a normalization factor
given by
$$ {\cal N} = \big[ 2 + 2 p_1p_2\cdots p_n \cos m \pi\big]^{-1/2}$$
where the quantities $p_i$, assumed to be reals,  stand for the
overlapping $\langle \Omega_i \vert \Omega'_i \rangle$ between two
single particle coherent states. It is important to stress that,
from an experimental point of view, superpositions of coherent
states are difficult to produce, and fundamentally this could be due
to extreme sensitivity to environmental decoherence. Experimental
efforts to create superpositions of coherent states was reported in
\cite{Brune}. Note also that evidence of such superpositions first
appeared in a study of a certain type of nonlinear Hamiltonian
evolution by \cite{Milburn1,Milburn2}, and the manifestation of
superpositions of coherent states was analyzed in detail by
\cite{Yurke1,Yurke2} (see also \cite{Buzek}).

\subsection{Pairwise partitioning and  qubit mapping}
To study the pairwise quantum correlations present in the
multipartite coherent state (\ref{multi-cs}), the whole system can
be partitioned in two different ways.
\subsubsection{Pure bipartite states}
We first consider the splitting of the entire system into two
subsystems; one subsystem containing any~$k$ $(1\le k\le n-1)$
particles and the other containing the remaining $n-k$ particles. In
this scheme, one writes the state (\ref{multi-cs}) as
\begin{equation}
\vert \Omega, \Omega', m, n \rangle = {\cal N} (\vert\Omega
\rangle_k \otimes \vert \Omega \rangle_{n-k} + e^{i m\pi } \vert
\Omega' \rangle_k \otimes \vert \Omega'
\rangle_{n-k})\label{partition1}
\end{equation}
where
$$ \vert \Omega \rangle_k =  \vert \Omega_1 \rangle\otimes
\vert \Omega_2 \rangle \otimes \cdots\otimes \vert \Omega_k \rangle
 \qquad  \vert \Omega' \rangle_k =  \vert \Omega'_1 \rangle\otimes
\vert \Omega'_2 \rangle \otimes \cdots\otimes \vert \Omega'_k
\rangle
$$
$$ \vert \Omega \rangle_{n-k} =  \vert \Omega_{k+1} \rangle\otimes
\vert \Omega_{k+2} \rangle \otimes \cdots\otimes \vert \Omega_n
\rangle
 \qquad  \vert \Omega' \rangle_{n-k} =  \vert \Omega'_{k+1}\rangle\otimes
\vert \Omega'_{k+2} \rangle \otimes \cdots\otimes \vert \Omega'_n
\rangle
$$
The whole system can be mapped into a pair of two logical qubits.
This can be done by introducing the orthogonal basis $\{ \vert {\bf
0} \rangle_k , \vert {\bf 1} \rangle_k\}$ defined as
\begin{equation}
\vert {\bf 0} \rangle_k = \frac{ \vert \Omega \rangle_k +  \vert
\Omega' \rangle_k}{\sqrt{2(1 + p_1p_2\cdots p_k)}}
   \qquad \vert {\bf
1} \rangle_k = \frac{\vert \Omega \rangle_k -  \vert \Omega'
\rangle_k}{{\sqrt{2(1- p_1p_2\cdots p_k)}}}\label{base1}
\end{equation}
for the first subsystem. Similarly, we introduce for the second
subsystem, containing the remaining $n-k$  particles, the orthogonal
basis $\{ \vert {\bf 0} \rangle_{n-k} , \vert {\bf 1}
\rangle_{n-k}\}$ given by
\begin{equation}
\vert {\bf 0} \rangle_{n-k} = \frac{ \vert \Omega \rangle_{n-k} +
\vert \Omega' \rangle_{n-k}}{\sqrt{2(1 + p_{k+1}p_{k+2} \cdots
p_{n})}}
   \qquad
   \vert {\bf
1} \rangle_{n-k} = \frac{\vert \Omega \rangle_{n-k} -  \vert \Omega'
\rangle_{n-k}}{{\sqrt{2(1- p_{k+1}p_{k+2} \cdots
p_{n}})}}\label{base2}
\end{equation}
Reporting the equations (\ref{base1}) and (\ref{base2}) in
(\ref{partition1}), one has the explicit form of the pure state
$\vert \Omega, \Omega', m, n \rangle$ in the basis $\{ \vert {\bf 0}
\rangle_{k} \otimes \vert {\bf 0} \rangle_{n-k} ,
 \vert {\bf 0} \rangle_{k} \otimes \vert {\bf 1} \rangle_{n-k} , \vert {\bf 1} \rangle_{k}
 \otimes \vert {\bf 0} \rangle_{n-k} , \vert {\bf 1} \rangle_{k} \otimes \vert {\bf 1}
 \rangle_{n-k}\}$. It is given by
\begin{equation}
\vert \Omega, \Omega', m, n \rangle = \sum_{\alpha= 0,1}
\sum_{\beta= 0,1} C_{\alpha,\beta} \vert \alpha \rangle_k \otimes
\vert \beta \rangle_{n-k}
\end{equation}
where
$$ C_{0,0} = {\cal N}(1 + e^{im\pi}) a_{k}a_{n-k}  , \qquad  C_{0,1} =  {\cal N} (1 -e^{im\pi}) a_{k}b_{n-k} $$
$$ C_{1,0} = {\cal N} (1 - e^{im\pi}) a_{n-k}b_{k}  , \qquad  C_{1,1} =  {\cal N} (1 + e^{im\pi}) b_{k}b_{n-k} $$
with
$$ a_k =  \sqrt{\frac{1+p_1p_2\cdots p_k}{2}}\qquad b_k = \sqrt{\frac{1- p_1p_2\cdots p_k}{2}}$$
$$ a_{n-k} =  \sqrt{\frac{1+p_{k+1}p_{k+2}\cdots p_n}{2}}\qquad b_{n-k} = \sqrt{\frac{1- p_{k+1}p_{k+2}\cdots p_n}{2}}$$
involving the overlapping $p_i = \langle \Omega_i \vert \Omega'_i
\rangle, ~ i = 1, 2, \cdots, n $.

\subsubsection{Mixed bipartite states}

The second  partitioning scheme can be realized by considering  the
bipartite reduced density matrix $\rho_{ij}$, which is obtained by
tracing out all other systems except subsystems or modes $i$ and
$j$. There are $n(n-1)/2$ different density matrices $\rho_{ij}$. It
is simply verified that the reduced density matrix describing the
subsystems $i$ and $j$ is
\begin{eqnarray}
\rho_{ij} = {\cal N}^2(\vert \Omega_i , \Omega_j \rangle \langle
\Omega_i , \Omega_j \vert +\vert \Omega'_i , \Omega'_j \rangle
\langle \Omega'_i , \Omega'_j | + e^{im\pi } q_{ij} |\Omega'_i ,
\Omega'_j  \rangle \langle \Omega_i , \Omega_j \vert +e^{-i m \pi
}q_{ij}\vert \Omega_i , \Omega_j \rangle \langle \Omega'_i ,
\Omega'_j \vert ). \label{rhoij-vect}
\end{eqnarray}
The quantity $q_{ij}$  occurring in (\ref{rhoij-vect}) is defined by
$$q_{ij} = p_1p_2\cdots \check{p}_i \cdots \check{p}_j \cdots p_n$$
where the notation $\check{p}_i$ and $\check{p}_j$  indicates that
$p_i$ and $p_j$ must be omitted from the product of the overlapping
of coherent states. Here also, one can map the reduced system into a
pair of two-qubits. In this sense, we introduce, for the subsystem
$i$, the orthogonal basis $\{\vert {\bf 0}_i\rangle ,\vert {\bf
1}_i\rangle \}$ defined such that
\begin{equation}
\vert\Omega_i\rangle \equiv  a_i \vert {\bf 0}_i \rangle + b_i \vert
{\bf 1}_i \rangle \qquad \vert \Omega'_i\rangle\equiv a_i \vert {\bf
0}_i \rangle - b_i
 \vert {\bf 1}_i \rangle~,\label{basei}
\end{equation}
where $$a_i = \sqrt{\frac{1+p_i}{2}} \qquad b_i =
\sqrt{\frac{1-p_i}{2}}.$$ Similarly for the subsystem $j$, we
introduce a second two dimensional orthogonal basis as
\begin{equation}
\vert\Omega_j\rangle \equiv  a_j \vert {\bf 0}_j \rangle + b_j \vert
{\bf 1}_j \rangle \qquad \vert \Omega'_j\rangle\equiv a_j \vert {\bf
0}_j \rangle - b_j
 \vert {\bf 1}_j \rangle~,\label{basej}
\end{equation}
where $$a_j = \sqrt{\frac{1+p_j}{2}} \qquad b_j =
\sqrt{\frac{1-p_j}{2}}.$$ Substituting Eqs.~(\ref{basei}) and
(\ref{basej}) into Eq.~(\ref{rhoij-vect}), we obtain the mixed
density matrix
\begin{equation}
\rho_{ij} = {\cal N}^2 \left( \begin{smallmatrix}
2a_i^2a_j^2(1+q_{ij}\cos m\pi)& 0
    & 0& 2a_ia_jb_ib_j(1+q_{ij}\cos m\pi
)\\
0  & 2a_i^2b_j^2(1-q_{ij}\cos m\pi ) & 2a_ia_jb_ib_j(1-q_{ij}\cos
m\pi
) & 0 \\
0  & 2a_ia_jb_ib_j(1-q_{ij}\cos m\pi ) & 2a_j^2b_i^2(1-q_{ij}\cos
m\pi
) & 0 \\
2a_ia_jb_ib_j(1+q_{ij}\cos m\pi ) & 0 & 0 & 2b_i^2b_j^2(1+q_{ij}\cos
m\pi)
\end{smallmatrix}
\right) \label{rhoij}
\end{equation}
in the basis $\{\vert{\bf 0}_i{\bf 0}_j\rangle ,\vert{\bf 0}_i{\bf
1}_j\rangle ,\vert{\bf 1}_i{\bf 0}_j\rangle ,
    \vert{\bf 1}_i{\bf 1}_j\rangle \}$.

\section{ Quantum discord}

\subsection{Geometric measure of quantum discord}
Evaluation of quantum discord, based on the original definition
given in \cite{Vedral-et-al,Ollivier-PRL88-2001}, involves a
difficult optimization procedure and analytical results were
obtained only in few cases
\cite{Luo-PRA77-2008,Dillenschneider-PRB78-2008,Sarandy-arXiv,Ali-qd}.
To overcome this difficulty Dakic et al introduced a geometric
measure of quantum discord \cite{Dakic2010}. It is defined as the
distance between a state $\rho$ of a bipartite system $AB$ and the
closest classical-quantum state presenting zero discord:
\begin{equation}
  D_{g}(\rho):=\min_{\chi}||\rho-\chi||^{2}
\end{equation}
where the minimum is over the set of zero-discord states $\chi$ and
the distance is the square norm in the Hilbert-Schmidt space. It is
given by
$$||\rho-\chi||^{2}:= {\rm Tr}(\rho-\chi)^2. $$
When the measurement is taken on the subsystem $A$, the zero-discord
state $\chi$ can be represented as \cite{Ollivier-PRL88-2001}
$$\chi= \sum_{i = 1,2}p_{i}|\psi_{i}\rangle\langle\psi_{i}|\otimes\rho_{i}
$$
where $p_i$ is a probability distribution, $\rho_{i}$ is the
marginal density matrix of $B$ and  $\{|\psi_1\rangle
,|\psi_2\rangle \}$ is an arbitrary orthonormal vector set. A
general two qubit state writes in Bloch representation as
\begin{eqnarray}
  \rho & = & \frac{1}{4}\left[ \sigma_{0}\otimes \sigma_{0} +\sum_{i}^{3}(x_{i}\sigma_{i}\otimes \sigma_{0}
   +y_{i} \sigma_{0}\otimes\sigma_{i})+\sum_{i,j=1}^{3}R_{ij}\sigma_{i}\otimes\sigma_{j}\right]
\end{eqnarray}
where $x_{i} = {\rm Tr}\rho(\sigma_{i}\otimes \sigma_{0}),~  y_{i} =
{\rm Tr}\rho(\sigma_{0}\otimes\sigma_{i})$ are components of local
Bloch vectors and $R_{ij} = {\rm
Tr}\rho(\sigma_{i}\otimes\sigma_{j})$ are components of the
correlation tensor. The operators $\sigma_i$ $( i = 1, 2, 3)$ stand
for the three Pauli matrices and $\sigma_0$ is the identity matrix.
The explicit expression of the geometric measure of quantum discord
is given by \cite{Dakic2010}:
\begin{equation}
  D_g(\rho)=\frac{1}{4}\left(||x||^{2}+||R||^{2}-k_{\rm {max}}\right) \label{eq:GMQD_original}
\end{equation}
where $x=(x_{1},x_{2},x_{3})^{T}$, $R$ is the matrix with elements
$R_{ij}$, and $k_{\rm{max}}$ is the largest eigenvalue of matrix
defined by
\begin{equation}
K := xx^{T}+RR^{T}. \label{matrix K}
\end{equation}
Denoting the eigenvalues of the $3\times 3$ matrix $K$ by
$\lambda_1$, $\lambda_2$ and $\lambda_3$ and considering
$||x||^{2}+||R||^{2}={\rm Tr}K$, we get an alternative compact form
of the geometric measure of quantum discord
\begin{equation}
D_g(\rho) = \frac{1}{4}~ {\rm min}\{ \lambda_1 + \lambda_2 ,
\lambda_1 + \lambda_3 , \lambda_2 + \lambda_3\}.\label{eq:GMQD_new}
\end{equation}
 This  form will be more convenient for our purpose.

\subsection{Geometric measure of quantum discord for pure coherent states}
Local unitary operations do not affect the quantum correlations
present in a quantum system. In this respect, the geometric quantum
discord in the pure state $\vert \Omega, \Omega', m, n \rangle$,
partitioned according the scheme (\ref{partition1}), can be
evaluated by making use of the Schmidt decomposition. Therefore, we
write the state $\vert \Omega, \Omega', m, n \rangle$ as
\begin{equation}
\vert \Omega, \Omega', m, n \rangle = \sqrt{\lambda_+} ~\vert +
\rangle_k \otimes \vert + \rangle_{n-k} + \sqrt{\lambda_-}~ \vert -
\rangle_k \otimes \vert - \rangle_{n-k} \label{shcmidt}
\end{equation}
where $\vert \pm \rangle_k$ denotes the eigenvectors of the reduced
density matrix $\rho_1$ associated with the first subsystem
containing $k$ particles. Similarly, $\vert \pm \rangle_{n-k}$
denotes the eigenvectors of the reduced density matrix $\rho_2$ for
the second subsystem.  The eigenvalues of the reduced density matrix
$\rho_1$ are given by
$$ \lambda_{\pm} = \frac{1}{2} \bigg(1 \pm \sqrt {1 - {\cal C}_{k,n-k}^2}\bigg)$$
in term of  the bipartite concurrence given by
\begin{equation}
{\cal C}_{k,n-k} = 2 \vert C_{0,0} C_{1,1} - C_{1,0}C_{0,1} \vert
=\frac{\sqrt{1-p_1^2p_2^2\cdots
p_k^2}\sqrt{1-p_{k+1}^2p_{k+2}^2\cdots p_n^2}}{1+p_1p_2\cdots
p_n\cos m\pi}.\label{Concurrence-pure}
\end{equation}
Note that the eigenvalues of the reduced matrix density $\rho_2$ are
identical to those of $\rho_1$. Separability and maximal
entanglement conditions are easily derivable from the equation
(\ref{Concurrence-pure}).

Using the prescription discussed in the previous subsection, the
derivation of  the analytical expression of the geometric quantum
discord is very simple. Indeed, one can verify that the matrix $K$
defined by  (\ref{matrix K}) writes
$$ K = {\rm diag} (4 \lambda_+ \lambda_-, 4 \lambda_+ \lambda_-, 2 ( \lambda_+^2 + \lambda_-^2) ).$$
Thus, using  the equation (\ref{eq:GMQD_new}),  one obtains the
bipartite geometric discord present in the state $\vert \Omega,
\Omega', m, n \rangle$ :
\begin{equation}
D_g(\vert \Omega, \Omega', m, n \rangle \langle \Omega, \Omega', m,
n  \vert)=\frac{1}{2} \frac{(1-p_1^2p_2^2\cdots
p_k^2)(1-p_{k+1}^2p_{k+2}^2\cdots p_n^2)}{(1+ p_1p_2\cdots p_n\cos
m\pi)^2}\label{discordpure}
\end{equation}
where we used the mapping defined by the equations (\ref{base1}) and
(\ref{base2}) to convert the whole system into  a pair of qubit
systems. It is remarkable that the result (\ref{discordpure}) can be
rewritten as
\begin{equation}
D_g(\vert \Omega, \Omega', m, n \rangle \langle \Omega, \Omega', m,
n  \vert) = \frac{1}{2}~{\cal C}_{k,n-k}^2,\label{qdpurestate}
\end{equation}
in term of the bipartite concurrence given by
(\ref{Concurrence-pure}). This  gives a very  simple relation
between the entanglement and the geometric quantum discord in a pure
multipartite coherent state. It must be noticed that for any
bipartite pure state, the quantum discord is exactly the
entanglement of formation. It follows that the result
(\ref{qdpurestate}) establishes a relation between the quantum
discord  and its geometrized version.

\subsection{Geometric measure of quantum discord for mixed states}

The bipartite mixed density $\rho_{ij}$ (\ref{rhoij}), obtained in
the second bipartition scheme, can be cast in the following compact
form
\begin{equation}
\rho_{ij} = \sum_{\alpha \beta} R_{\alpha \beta}
\sigma_{\alpha}\otimes \sigma_{\beta}
\end{equation}
where the non vanishing matrix elements $R_{\alpha \beta}$ $(\alpha,
\beta = 0,1,2,3)$ are given by
$$ R_{00} = 1, \quad R_{11} = 2{\cal N}^2 \sqrt{(1- p_i^2)(1- p_j^2)}, \quad R_{22} = -2{\cal N}^2 \sqrt{(1- p_i^2)(1- p_j^2)}~q_{ij}\cos
m\pi,$$ $$ R_{33} = 2{\cal N}^2 (p_ip_j + q_{ij}\cos m\pi), \quad
R_{03} = 2{\cal N}^2 (p_j + p_i q_{ij}\cos m\pi), \quad R_{30} =
2{\cal N}^2 (p_i + p_j q_{ij}\cos m\pi).$$ Having mapped the
bipartite system $\rho_{ij}$ into a pair of two qubits (see
equations (\ref{basei}) and (\ref{basej})), now we use it to
investigate the pairwise geometric quantum discord according the
prescription presented in the subsection 3.1. It is easy, modulo
some obvious substitutions, to check that the eigenvalues of the
matrix $K$, defined in (\ref{matrix K}), are given by
\begin{equation}
\lambda_1 = 4 {\cal N}^4 \bigg[(1 + p_i^2)(p_j^2 + q_{ij}^2) + 4
(p_1p_2\cdots p_n) \cos m\pi\bigg]\label{lambda1}
\end{equation}
\begin{equation}
\lambda_2 = 4 {\cal N}^4 (1 - p_i^2)(1 - p_j^2)\label{lambda2}
\end{equation}
\begin{equation}
\lambda_3 = 4 {\cal N}^4 (1 - p_i^2)(1 -
p_j^2)q_{ij}^2\label{lambda3}
\end{equation}
Clearly, we have $\lambda_3 < \lambda_2$. Thus,  the equation
(\ref{eq:GMQD_new}) gives
\begin{equation}
D_g = \frac{1}{4} {\rm min}\{  \lambda_1 + \lambda_3 , \lambda_2 +
\lambda_3\}.\label{Dgplus}
\end{equation}
For mixed states $\rho_{ij}$, the explicit expression of geometric
quantum discord is
\begin{equation}
D_g =  \frac{1}{4} \frac{(1 - p_i^2)(1 - p_j^2)(1+q_{ij}^2)}{(1 +
p_1p_2\cdots p_n\cos m\pi)^2}\label{Dgplus-general}
\end{equation}
 when the condition
$\lambda_1 > \lambda_2$ is satisfied or
\begin{equation}
D_g = \frac{1}{4}\frac{(1 + p_i^2)(p_j^2 + q_{ij}^2)+ (1 - p_i^2)(1
- p_j^2)q_{ij}^2 + 4 (p_1p_2\cdots p_n) \cos m\pi}{(1 + p_1p_2\cdots
p_n\cos m\pi)^2} \label{Dgmoins-general}
\end{equation}
in the situation where $ \lambda_1 < \lambda_2$. It is interesting
to note that in the particular case where the system contains only
two particles (i.e., $n=2$), we have $q_{12} = 1$ and one can verify
that $\lambda_1 > \lambda_2$ so that the equation
(\ref{Dgplus-general}) reduces to
$$D_g =  \frac{1}{2} \frac{(1 -
p_1^2)(1 - p_2^2)}{(1 + p_1p_2\cos m\pi)^2}$$ which coincides the
quantum discord (\ref{discordpure}) when $n=2$ and $k=1$. The
results obtained in this section provide us with the explicit
expressions of bipartite geometric quantum discord involving
nonorthogonal (pure as well as mixed) states . An illustration of
these results  will be considered in Section 5 for some specific
superpositions of multipartite coherent states. But before to do
this, we shall, in the following section, discuss  the dynamical
evolution of pairwise quantum correlations present in the state
$\vert \Omega, \Omega', m, n \rangle$.

\section{Evolution of quantum correlations under dephasing channel}

It has been observed that a pair of entangled qubits, interacting
with noisy environments, becomes separable  in a finite time
\cite{Yu}. The phenomenon of total loss of entanglement, termed in
the literature "entanglement sudden death", was experimentally
confirmed \cite{Almeida}. The entanglement sudden death depends on
the nature of the system-environment interaction. Various
decoherence channels, Markovian as well non Markovian, were
investigated. In other hand, it has been shown that under some
specific channels, when entanglement die in a finite time, the
quantum discord vanishes only in asymptotic time
\cite{Werlang-PRA-2009}. This reflects that the quantum discord is
more robust against to decoherence than entanglement.

In this section, we investigate the dynamics of bipartite quantum
correlations (entanglement and quantum discord) of the multipartite
coherent states $\vert \Omega, \Omega' , m, n\rangle$ under a
dephasing dissipative channel. We use the Kraus operator approach
 which describes conveniently the dynamics of two
qubits interacting independently with individual environments (for
more details, see for instance \cite{NC-QIQC-2000}). The time
evolution of the bipartite density $\rho$ can be written compactly
as
$$ \rho(t) = \sum_{\mu, \nu}E_{\mu , \nu}(t) ~\rho(0)~ E_{\mu , \nu}^{\dagger}(t) $$
where the so-called Kraus operators
$$E_{\mu , \nu}(t) = E_{\mu}(t)\otimes E_{\nu}(t)\quad {\rm such ~that} \quad \sum_{\mu,
\nu}E_{\mu , \nu}^{\dagger}  E_{\mu , \nu} = \mathbb{I}$$ are the
tensorial product of the operators $E_{\mu}$ describing the
one-qubit quantum channel effects. The non-zero Kraus operators for
a dephasing channel are given by \cite{Werlang-PRA-2009}
\begin{equation}
E_0 = {\rm diag}( 1 , \sqrt{1 - \gamma})  \qquad  E_1 = {\rm diag}(
0 , \sqrt{\gamma}) \label{kraus}
\end{equation}
with $ \gamma = 1 - e^{-\Gamma t}$ and $\Gamma$ denoting the decay
rate.

We first consider the temporal evolution of geometric quantum
discord for the pure state given by (\ref{partition1}).  Using the
definition of the dephasing channel the Kraus description
(\ref{kraus}) and the Schmidt decomposition given by
(\ref{shcmidt}), it is simply verified that the matrix $K$ defined
by (\ref{matrix K}) becomes
$$K(t) = {\rm diag} (4 e^{-2\Gamma t}\lambda_+ \lambda_-, 4 e^{-2\Gamma t} \lambda_+ \lambda_-, 2 ( \lambda_+^2 + \lambda_-^2) )$$
and the geometric discord is then given by
$$ D_g(t)= 2~ e^{-2\Gamma t}~ \lambda_+ \lambda_-.$$
Under the dephasing channel, the concurrence evolves as
$${\cal C}_{k,n-k}(t) = e^{-2\Gamma t}~ {\cal C}_{k,n-k}$$ where ${\cal C}_{k,n-k}$ is the
concurrence given by (\ref{Concurrence-pure}) and we have
$$ D_g(t) = \frac{1}{2}~{\cal C}_{k,n-k}^2(t).$$
Clearly, for pure states (initially entangled) geometric discord as
well as entanglement
 vanishes only in the asymptotic time limit. This dissipative channel induces an exponential decay of the
geometric discord between the two sub-components of the system. This
situation
 becomes completely different for mixed density evolving under a
 dephasing channel. Indeed, it is easy  to check that the density matrix (\ref{rhoij}) evolves as
\begin{equation}
\rho_{ij} (t) = {\cal N}^2 \left( \begin{smallmatrix}
2a_i^2a_j^2(1+q_{ij}\cos m\pi)& 0
    & 0& 2(1 -
\gamma)a_ia_jb_ib_j(1+q_{ij}\cos m\pi
)\\
0  & 2a_i^2b_j^2(1-q_{ij}\cos m\pi ) & 2(1 -
\gamma)a_ia_jb_ib_j(1-q_{ij}\cos m\pi
) & 0 \\
0  & 2(1 - \gamma)a_ia_jb_ib_j(1-q_{ij}\cos m\pi ) &
2a_j^2b_i^2(1-q_{ij}\cos m\pi
) & 0 \\
2(1 - \gamma)a_ia_jb_ib_j(1+q_{ij}\cos m\pi ) & 0 & 0 &
2b_i^2b_j^2(1+q_{ij}\cos m\pi)
\end{smallmatrix}
\right), \label{rhoij(t)}
\end{equation}
which can be written in the Bloch representation as follows
\begin{equation}
\rho_{ij} (t) = \sum_{\alpha \beta} R_{\alpha \beta}(t)
\sigma_{\alpha} \otimes \sigma_{\beta}
\end{equation}
where all correlation matrix elements are time independent except
$R_{11}(t)$ and $R_{22}(t)$ which become
$$R_{11}(t) = e^{-\Gamma t} R_{11} \qquad R_{22}(t) = e^{-\Gamma t} R_{22}.$$
We will employ the concurrence as a measure of bipartite
entanglement for the state (\ref{rhoij(t)}). We recall that for
$\rho_{AB}$ the density matrix for a pair of qubits $A$ and $B$, the
concurrence is~\cite{Hil97}
\begin{equation}
C_{AB}=\max \left\{ c_1- c_2 - c_3 - c_4,0\right\} \label{eq:c1}
\end{equation}
for~$c_1\ge c_2\ge c_3\ge c_4$  the square roots of the eigenvalues
of the "spin-flipped" density matrix
\begin{equation}
\varrho_{AB}\equiv\rho_{AB}(\sigma_y\otimes\sigma
_y)\rho_{AB}^*(\sigma_y\otimes \sigma_y) \label{eq:c2}
\end{equation}
where the star stands for complex conjugation and $\sigma_y \equiv
\sigma_2$ is the usual Pauli matrix. Thus, after some algebra, one
shows that the entanglement in the state $\rho_{ij}$ is quantified
by the concurrence
$$ C_{ij}(t) = 2~ {\rm max} \{ 0 , \Lambda_1(t) , \Lambda_2(t) \}$$
where
$$ \Lambda_1(t) = \frac{1}{2} {\cal N}^2  \sqrt{(1 - p_i^2)(1 - p_j^2)} \bigg[   (1 - \gamma)(1+q_{ij}\cos m\pi ) - (1-q_{ij}\cos m\pi )\bigg]$$
$$ \Lambda_2(t) = \frac{1}{2} {\cal N}^2  \sqrt{(1 - p_i^2)(1 - p_j^2)} \bigg[   (1 - \gamma)(1-q_{ij}\cos m\pi ) - (1+q_{ij}\cos m\pi )\bigg].$$
It follows that the concurrence is given by
$$ C_{ij} (t)= \frac{1}{2}~ \frac{\sqrt{(1 - p_i^2)(1 - p_j^2)}}{1 + p_1p_2\cdots p_n\cos m\pi}\bigg[ e^{-\Gamma t}(1 + q_{ij}) - (1-q_{ij})\bigg]$$
for
$$ t < t_0 = \frac{1}{\Gamma} [ ~\ln(1 + q_{ij}) -  \ln(1 - q_{ij})]$$
and the system is entangled. However, for $ t \ge t_0$, the
concurrence is zero and the entanglement disappears., i.e. the
system is separable. This results shows that under the dephasing
channel, the entanglement suddenly vanishes at $t = t_0$. Note that
the bipartite system under consideration is in general entangled in
the absence of external interaction. Indeed, for $ t = 0$, the
concurrence is given by
$$ C_{ij}(0)=  \frac{q_{ij}\sqrt{(1 - p_i^2)(1 - p_j^2)}}{1 + p_1p_2\cdots p_n\cos m\pi},$$
and vanishes only in the special cases $q_{ij} = 0$ or $ p_i = 1$ or
$p_j = 1$. Next, we investigate the temporal evolution of the
quantum discord. As above, to study the dephasing channel effect, we
compute the evolution of the matrix $K$ after dephasing interaction.
We obtain a diagonal matrix and the corresponding elements are
$$ \lambda_1(t) =  R_{03}^2 + R_{33}^2 = \lambda_1$$
$$ \lambda_2(t) =  R_{11}^2(t) =  e^{-2\Gamma t} \lambda_2$$
$$ \lambda_3(t) =  R_{22}^2 (t)= e^{-2\Gamma t} \lambda_3$$
where $\lambda_1$ , $\lambda_2$ and $\lambda_2$ are given by
(\ref{lambda1}), (\ref{lambda2}) and (\ref{lambda3}) respectively.

The mixed states $\rho_{ij}(t)$ are purely classical (zero quantum
discord) if and only if $p_i \longrightarrow 0$ for $i = 1, 2,
\cdots, n$. To show this result, we used the criteria provided in
\cite{Dakic2010} to determine states with vanishing quantum discord.
This criteria uses the rank of the correlation matrix (the number of
non zero eigenvalues)  as a simple discord witness and states that
if the rank of the correlation tensor is greater than 2 (for
qubits), the state has non vanishing quantum discord. It is
important to note that in the  limiting case $p_i \rightarrow 0 $,
the two states $|\Omega_i \rangle $ and $|\Omega'_i\rangle $
approach orthogonality and an orthogonal basis can be constructed
such that $\vert {\bf 0}\rangle\equiv \vert \Omega_i \rangle$ and
$\vert{\bf 1}\rangle \equiv \vert \Omega'_i \rangle$. In this limit,
the state $ \vert \Omega,\Omega', m ,n \rangle$ reduces to the
multipartite ${\rm GHZ}$ state
\begin{equation}
 \vert\text{GHZ}\rangle_n = \frac
1{\sqrt{2}}(\vert {\bf 0}\rangle \otimes |{\bf 0}\rangle \otimes
        \cdots \otimes\vert {\bf 0}\rangle
    +e^{i m \pi}\vert {\bf 1}\rangle \otimes
    \vert {\bf 1}\rangle \otimes \cdots \otimes
\vert {\bf 1}\rangle).
\end{equation}

In summary, the dynamic evolutions of  geometric quantum discord and
entanglement for mixed states,  under  a dephasing channel, show
different behaviors. The life time of pairwise quantum discord, in
multipartite coherent states, is infinite while the entanglement
suddenly disappears after a finite time of interaction in bipartite
mixed states. This agrees well with the results recently obtained by
Ferraro et al \cite{Ferraro} (see also \cite{ Cole}) showing that
the interaction of a quantum system with a dissipative environment
can't induce a sudden death of quantum discord contrarily to
entanglement which can disappear suddenly and permanently.

\section{ Some special cases}
 To illustrate the
results of the previous sections, we shall focus on the coherent
states associated with Weyl-Heisenberg, $SU(2)$ and $SU(1,1)$
symmetries. They are labeled by a a single complex variable $z$.
Also, to simplify further our illustration a special form of
(\ref{multi-cs}) is considered. This is given by
\begin{equation}
 \vert z, m, n \rangle ={\cal N}(\vert z
\rangle\otimes \vert z \rangle\otimes \cdots\otimes \vert z \rangle
+e^{im\pi}|- z \rangle\otimes |- z \rangle\otimes \cdots\otimes |- z
\rangle) \label{multi-cs-z}
\end{equation}
where here again $m \in \mathbb{Z}$ and the normalization factor
${\cal N}$ rewrites
$$ {\cal N} = \big[ 2 + 2 p^n \cos m \pi\big]^{-1/2}.$$
The quantity $p$ is the overlapping $\langle z \vert - z \rangle$
between two single particle coherent states of the same amplitude
and opposite phase.  It is given by
$$ \langle z \vert - z \rangle = \exp(- 2 \vert z \vert^2)$$
for Glauber or Weyl-Heisenberg coherent states $( z \in
\mathbb{C})$. For $su(1,1)$ algebra, this kernel is given
$$\langle z \vert - z \rangle =  \bigg( \frac{1 - \vert z \vert^2}{1 + \vert z \vert^2} \bigg)^{2k}$$
for group theory or Perelomov coherent states ( $\vert z \vert < 1$
and $k$ is the Chen parameter characterizing the discrete-series
representations of $SU(1,1)$ Lie group). For $j$-spin coherent
states or $su(2)$ coherent states, the quantity $\langle z \vert - z
\rangle$ is
$$ \langle z \vert - z \rangle = \bigg( \frac{1 - \vert z \vert^2}{1 + \vert z \vert^2} \bigg)^{2j}$$
with $ z \in \mathbb{C}$. More details concerning coherent states
theory can be found in the references \cite{Perelomov,Ali,Gazeau}.

\noindent For the pure case obtained in the partitioning scheme
defined by (\ref{partition1}), the equation (\ref{discordpure})
gives
\begin{equation}
D_g(\vert z, m, n \rangle \langle  z, m, n \vert) = \frac{1}{2}
\frac{(1 - p^{2k})(1 - p^{2(n-k)})}{(1 + p^n \cos
m\pi)^2}\label{qdpurestate-special}
\end{equation}

\noindent In the second partitioning scheme for the state $\vert z
,m ,n\rangle$ given by (\ref{rhoij}), all the reduced density
matrices $\rho_{ij}$ are identical ($p_i = p$ for $i = 1, 2, \cdots,
n$). In this special case, it is simple to see that the eigenvalues
of the matrix $K$ given (\ref{lambda1}), (\ref{lambda2}) and
(\ref{lambda3})  take the following forms
\begin{equation}
 \lambda_1 =  \frac{(p^2 + p^{2(n-2)})(1+p^2) + 4p^n\cos m\pi}{(1 +
p^n \cos m\pi)^2},\label{lambda1-part}
\end{equation}
\begin{equation}
 \lambda_2 =  \frac{(1 - p^2)^2}{(1 + p^n \cos m\pi)^2},\label{lambda2-part}
\end{equation}
\begin{equation}
 \lambda_3 =  \frac{ p^{2(n-2)}(1 - p^2)^2}{(1 + p^n \cos m\pi)^2},\label{lambda3-part}
\end{equation}
 and the geometric quantum discord for the mixed state $\rho_{12}$ is
\begin{equation}
D_g = \frac{1}{4} (\lambda_2 + \lambda_3) = \frac{1}{4}  \frac{(1+
p^{2(n-2)})(1 - p^2)^2}{(1 + p^n \cos m\pi)^2}\label{Dgplus}
\end{equation}
when
\begin{equation}
(p^2 + 1)(1+p^{n-2}\cos m\pi) - 2(1 - p^2) \ge
0,\label{conditionplus}
\end{equation}
or
\begin{equation}
D_g = \frac{1}{4} (\lambda_1 + \lambda_3) = \frac{1}{4} \frac{(p^2 +
p^{2(n-2)})(1+p^2) + 4p^n\cos m\pi + p^{2(n-2)}(1 - p^2)^2}{(1 + p^n
\cos m\pi)^2}\label{Dgmoins}
\end{equation}
when
\begin{equation}
(p^2 + 1)(1+p^{n-2}\cos m\pi) - 2(1 - p^2) \le
0.\label{conditionmoins}
\end{equation}
The results (\ref{Dgplus}) and (\ref{Dgmoins}) follow from the
expressions (\ref{Dgplus-general}) and (\ref{Dgmoins-general}),
respectively. The limiting case $p \longrightarrow 1$ must be
treated carefully for antisymmetric states ($m$ odd).  In this
limit, the eigenvalues of the matrix $K$ (\ref{matrix K}) given by
(\ref{lambda1-part}), (\ref{lambda2-part}) and (\ref{lambda3-part})
reduce to
$$\lambda_1 = \bigg(1- \frac{4}{n}\bigg)^2 + \bigg(1- \frac{2}{n}\bigg)^2
\qquad \lambda_2 = \lambda_3 = \frac{4}{n^2}, $$ and the geometric
quantum discord is
$$ D_g = \frac{2}{n^2}.$$
It is interesting to note that when $p \longrightarrow 1$ ($z
\longrightarrow 0 $), the state (\ref{multi-cs-z}) with $m$ odd
reduces to a superposition similar to the so-called Werner state
\cite{Werner}. Indeed, when the  state $\vert z , m = 1, n \rangle$
involves Glauber coherent states,  one can verify that
\begin{equation}
\vert z \longrightarrow 0 , m = 1, n \rangle \sim \vert\text{\rm
W}\rangle_n
    = \frac{1}{\sqrt{n}}(\vert 1\rangle \otimes\vert 0 \rangle \otimes \cdots\otimes
       \vert0\rangle  +\vert 0\rangle \otimes\vert 1\rangle \otimes\ldots\otimes \vert0\rangle
      +\cdots
   + \vert 0\rangle \otimes\vert 0\rangle  \otimes \cdots\otimes \vert 1\rangle)~.
\label{Wstate}
\end{equation}
Here $\vert n\rangle $ $(n=0,1)$ denote the usual harmonic
oscillator Fock states. A similar superposition is obtained  for the
antisymmetric states $\vert z , m = 1 ({\rm mod}~ 2), n \rangle$
involving $SU(2)$ and $SU(1,1)$ coherent states when $z
\longrightarrow 0 $. This can be done using the contraction
procedure to pass from $SU(2)$ and $SU(1,1)$ to Weyl-Heisenberg
algebra.

In the special case $n = 2$, the equations (\ref{Dgplus}),
(\ref{conditionplus}), (\ref{Dgmoins}) and (\ref{conditionmoins})
give
\begin{equation}
D_g = \frac{1}{2} \frac{(1-p^2)^2}{(1+p^2\cos m\pi)^2}.
\end{equation}
This expression coincides with the geometric discord given by
(\ref{qdpurestate-special}) for $n=2$. Indeed,  for $n = 2$ the
density $\rho_{12}$ is pure. It must be noticed that for $m$ even,
the maximum  of the geometric quantum discord is $1/2$ which is
reached in the orthogonal limiting case  $p \longrightarrow  0$ (see
the figure 1). However, for $m$ odd, the geometric quantum discord
takes the constant value $1/2$.

The situation is slightly different for $n = 3$. For $m$ even, the
condition (\ref{conditionplus}) (resp. (\ref{conditionmoins})) is
satisfied when $ \sqrt{2} - 1 \leq p \leq 1$ (resp. $ 0 \leq p \leq
\sqrt{2} - 1$). It follows that the geometric quantum discord is
given by
$$ D_g =  \frac{1}{4} \frac{p^2(1+p)^2(2 + (1-p)^2)}{(1+p^3)^2} $$
for $ 0 \leq p \leq \sqrt{2} - 1$ and
$$ D_g =  \frac{1}{4} \frac{(1-p^2)^2(1+p^2)}{(1+p^3)^2} $$
for $ \sqrt{2} - 1 \leq p \leq 1$. However for antisymmetric states
(i.e. $m$ odd), the condition (\ref{conditionmoins}) is satisfied
for $ 0 \leq p < 1$ and the quantum discord reads as
$$ D_g =  \frac{1}{4} \frac{p^2(1-p)^2(2 + (1+p)^2)}{(1-p^3)^2}. $$

The behavior of geometric quantum discord for mixed states with
$n>2$ is given in the figures 2 and 3. Figure 2 gives a plot of
geometric quantum discord versus the overlapping $p$ for symmetric
multipartite coherent states ($m$ even). As seen from the figure,
after an initial increasing, the quantum discord decreases to vanish
when $p = 1$. The maximum value of quantum discord occurs when
$\lambda_1$ (\ref{lambda1-part}) and $\lambda_2$
(\ref{lambda2-part}) coincide. This maximum decreases as the
particle number $n$ increases. In figure 3, we give a plot of the
geometric quantum discord for $m$ odd (the antisymmetric case) and
different values of $n$. We have already noticed that in the limit
$p \longrightarrow 1$, we obtain a $\vert W \rangle_n$ state
(\ref{Wstate}) and the pairwise quantum discord behaves as $n^{-2}$
and vanishes only in the limit of large number of particles. Note
also that for $ n = 3$ and $n = 4$, the geometric quantum discord
increases to reach its maximal value in the limit $p \longrightarrow
1$. However, for $ n \ge 5$, the maximal value of quantum discord is
larger than the one obtained in the limit $p \longrightarrow 1$.
More interestingly, the quantum discord starts increasing  to reach
its maximal value for some fixed $p \neq 1$ and decreases after to
coincide with the quantum discord characterizing a  Werner state
obtained for $p \longrightarrow 1$. It is remarkable that for
antisymmetric quantum states containing more than five particles,
the maximal value of quantum discord increases as $n$ increases
contrarily to the symmetric states (see figure 2) for which the
corresponding maximal value diminishes as $n$ takes large values.

\begin{center}
  \includegraphics[width=4in]{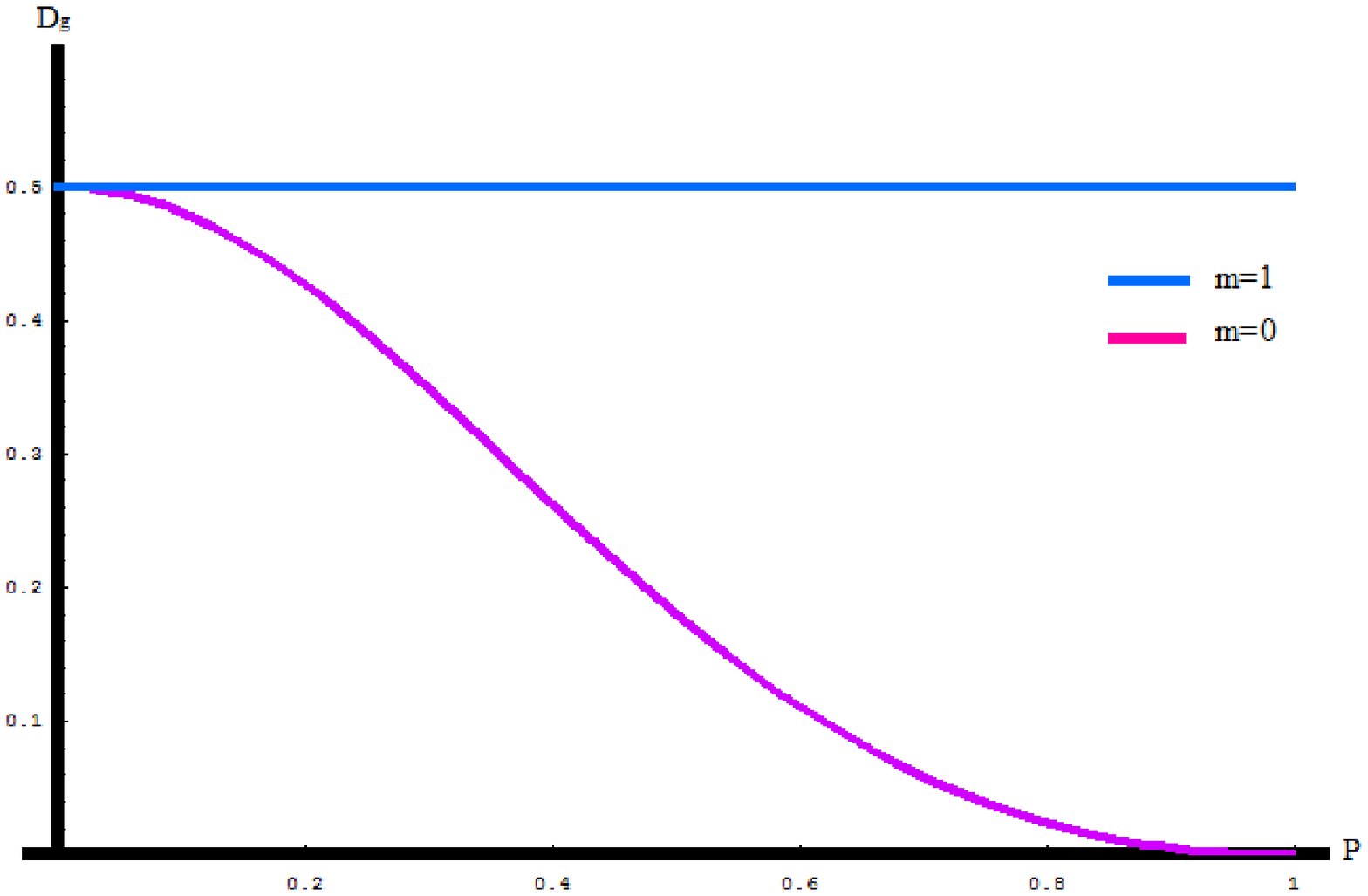}\\
{\bf Figure 1.}  {\sf The pairwise geometric quantum discord $D_g$
versus the overlapping $p$ for  $n=2$.}
\end{center}

\begin{center}
  \includegraphics[width=4in]{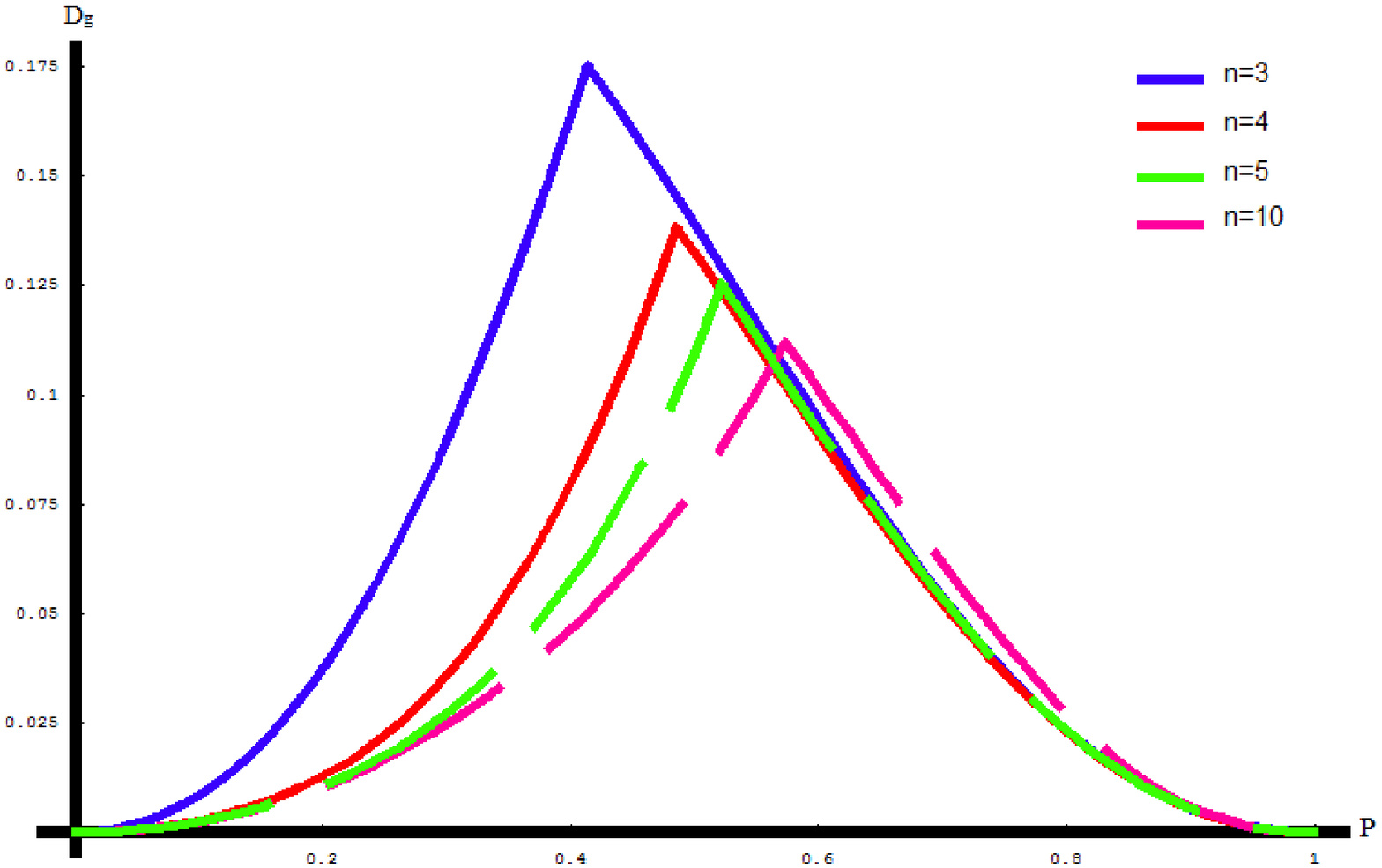}\\
{\bf Figure 2.}  {\sf The pairwise geometric quantum discord $D_g$
versus the overlapping $p$ for symmetric states.}
\end{center}
\begin{center}
  \includegraphics[width=4in]{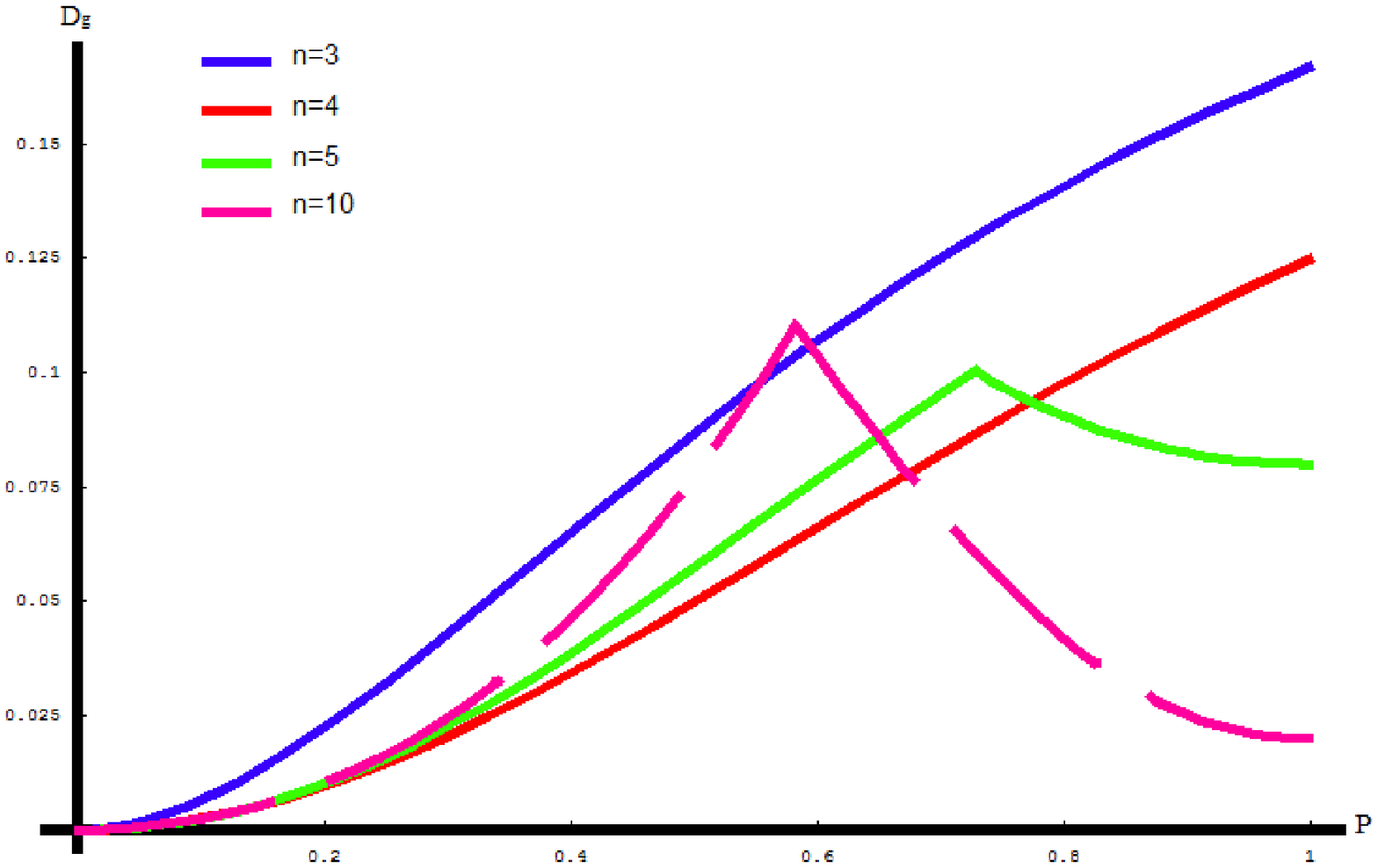}\\
{\bf Figure 3.} {\sf The pairwise geometric quantum discord $D_g$
versus the overlapping $p$ for anti-symmetric states}
\end{center}
\section{ Summary and concluding remarks}
In this paper, we have studied the pairwise quantum correlations in
 multipartite coherent states. A special attention was paid to
bipartite quantum discord. We used the geometric definition of the
quantum discord introduced in \cite{Dakic2010}. We have derived the
explicit expressions of this kind of correlation. We also
investigated the bipartite entanglement. We used two inequivalent
schemes in partitioning the system containing $n$ modes. The first
one consists in splitting the whole system in two parts: one
containing $k$ particles and the second is made of the remaining
$n-k$ particles leading to a pure state system. The second
bi-partitioning scheme is realized by tracing out $n-2$ particles or
modes and gives a mixed state. For each scheme, we mapped the system
in a pair of two qubits.  This mapping is helpful in investigating
the bipartite quantum correlation in a quantum state involving
coherent states which constitute the perfect example of non
orthogonal states. In particular the obtained quantum correlations
(quantum discord as well as entanglement) for antisymmetric
superpositions can be viewed as interpolating between ones present
in Greenberger-Horne-Zeilinger ($\vert {\rm GHZ} \rangle_n$) and
Werner ($\vert {\rm W} \rangle_n$) states. The analytic expressions
for geometric quantum discord are corroborated by some numerical
analysis. For a pure state (the first partitioning scheme), the
geometric quantum discord is proportional to the concurrence.
However, for a mixed bipartite state (the second partitioning
scheme), the two concepts are completely different. To show that the
quantum discord is a kind of correlation beyond the entanglement, we
have studied the dynamical evolution of entanglement under a very
simple noisy channel (dephasing channel). Indeed, for multipartite
coherent states, the pairwise entanglement disappears completely
after a finite time interaction while the quantum discord  is more
resilient. Finally, as prolongation of the present work, it will be
interesting  to compare the geometric quantum discord and the
quantum discord as defined in
\cite{Vedral-et-al,Ollivier-PRL88-2001}. Also, the method, discussed
in this work to investigate the bipartite correlations , can be
extended to examine the multipartite quantum correlations in
generalized coherent states in the spirit of the analysis recently
proposed in \cite{Okrasa}.

\end{document}